# Dispersion in disks[*]

Adrian Dumitrescu[†]    Minghui Jiang[‡]

November 1, 2018

**Abstract**

We present three new approximation algorithms with improved constant ratios for selecting $n$ points in $n$ disks such that the minimum pairwise distance among the points is maximized.

(1) A very simple $O(n \log n)$-time algorithm with ratio 0.511 for disjoint unit disks.

(2) An LP-based algorithm with ratio 0.707 for disjoint disks of arbitrary radii that uses a linear number of variables and constraints, and runs in polynomial time.

(3) A hybrid algorithm with ratio either 0.4487 or 0.4674 for (not necessarily disjoint) unit disks that uses an algorithm of Cabello in combination with either the simple $O(n \log n)$-time algorithm or the LP-based algorithm.

The LP algorithm can be extended for disjoint balls of arbitrary radii in $\mathbb{R}^d$, for any (fixed) dimension $d$, while preserving the features of the planar algorithm. The algorithm introduces a novel technique which combines linear programming and projections for approximating Euclidean distances. The previous best approximation ratio for dispersion in disjoint disks, even when all disks have the same radius, was 1/2. Our results give a partial answer to an open question raised by Cabello, who asked whether the ratio 1/2 could be improved.

## 1 Introduction

Let $\mathcal{R}$ be a family of $n$ subsets of a metric space. The problem of *dispersion in* $\mathcal{R}$ is that of selecting $n$ points, one in each subset, such that the minimum inter-point distance is maximized. This dispersion problem was introduced by Fiala et al. [8] as "systems of distant representatives", generalizing the classic problem "systems of distinct representatives". An especially interesting version of the dispersion problem, which has natural applications to wireless networking and map labeling, is in a geometric setting where $\mathcal{R}$ is a set of disks in the plane.

Let $\mathcal{D}$ be a set of $n$ disks in the plane. Dispersion in disks is the problem of selecting $n$ points, one from each disk in $\mathcal{D}$, such that the minimum pairwise distance of the selected points is maximized. Dispersion in disks is a hard problem. Fiala et al. [8] showed that dispersion in unit disks is already NP-hard. It is not difficult to modify their construction, which gives a reduction from Planar-3SAT, to show that dispersion in disjoint unit disks is also NP-hard. Moreover, by a slackness argument [9, 10], the same construction also implies that the problem is APX-hard; i.e, unless P = NP, the problem does not admit any polynomial-time approximation scheme. On the positive side, Cabello [3] presented an approximation algorithm PLACEMENT with ratio 3/8 for dispersion in arbitrary disks, and another approximation algorithm CENTERS with ratio 1/2 for dispersion in disjoint disks. Cabello also showed that applying both algorithms PLACEMENT and CENTERS and taking the better solution yields an approximation algorithm with ratio about 0.4465 (1/2.2393) for dispersion in unit disks.

---

[*]A preliminary version of this paper [5] appeared in the Proceedings of the 27th International Symposium on Theoretical Aspects of Computer Science, Nancy, France, March 2010.

[†]Department of Computer Science, University of Wisconsin–Milwaukee, WI 53201-0784, USA. Email: dumitres@uwm.edu. Supported in part by NSF CAREER grant CCF-0444188. Part of the research by this author was done at Ecole Polytechnique Fédérale de Lausanne.

[‡]Department of Computer Science, Utah State University, Logan, UT 84322-4205, USA. Email: mjiang@cc.usu.edu. Supported in part by NSF grant DBI-0743670.

We first introduce some preliminaries. For two points $p = (x_p, y_p)$ and $q = (x_q, y_q)$ in the plane, let $|pq|$ denote the Euclidean distance between them: $|pq| = \sqrt{(x_p - x_q)^2 + (y_p - y_q)^2}$. A *unit disk* is a disk of radius one. The *distance between two disks* is the distance between their centers; e.g., the distance between two tangent disks of radii $r_1$ and $r_2$ is $r_1 + r_2$. Throughout this paper, *disjoint* means *interior-disjoint*.

We now review the previous best $\frac{1}{2}$-approximation for dispersion in disjoint disks, which is achieved by a naive algorithm CENTERS that simply selects the centers of the given disks as the points [3]. Let $\mathcal{D} = \{\Omega_1, \ldots, \Omega_n\}$ be a set of $n$ disjoint disks of arbitrary radii in the plane. For each $i$, let $r_i$ be the radius of $\Omega_i$. For $i \neq j$, let $\delta_{ij}$ be the distance between $\Omega_i$ and $\Omega_j$. Let $\delta$ be the minimum pairwise distance of the disks in $\mathcal{D}$, i.e., $\delta = \min_{i \neq j} \delta_{ij}$. Let OPT denote an optimal solution and CEN denote the solution returned by CENTERS. We clearly have CEN $= \delta$. Since the disks are disjoint,
$$r_i + r_j \leq \delta_{ij}, \ i \neq j.$$
It follows that
$$\text{OPT} \leq \min_{i \neq j}(\delta_{ij} + r_i + r_j) \leq 2 \min_{i \neq j} \delta_{ij} = 2\delta. \tag{1}$$
Consequently, the algorithm CENTERS achieves an approximation ratio of
$$\frac{\text{CEN}}{\text{OPT}} \geq \frac{\delta}{2\delta} = \frac{1}{2}$$
for disjoint disks of arbitrary radii. Cabello asked whether this trivial $\frac{1}{2}$-approximation can be improved, even when all disks are not only disjoint but also have the same radius [3, p. 72].

We start with a very simple and efficient algorithm that achieves a ratio better than 1/2 for dispersion in disjoint unit disks.

**Theorem 1.** *There is an $O(n \log n)$-time approximation algorithm with ratio $0.511$ for dispersion in $n$ disjoint unit disks.*

Using linear programming, we then obtain the following substantially better approximation for dispersion in disjoint disks of arbitrary radii.

**Theorem 2.** *There is an LP-based approximation algorithm, with $O(n)$ variables and constraints, and running in polynomial time, that achieves approximation ratio $0.707$, for dispersion in $n$ disjoint disks of arbitrary radii. Moreover, the algorithm can be extended for disjoint balls of arbitrary radii in $\mathbb{R}^d$, for any (fixed) dimension $d$, while preserving the same features.*

We next improve the 0.4465-approximation for dispersion in (not necessarily disjoint) unit disks by a hybrid algorithm that uses the algorithm PLACEMENT of Cabello in combination with either the simple $O(n \log n)$-time algorithm in Theorem 1 or the LP-based algorithm in Theorem 2.

**Theorem 3.** *In combination with an algorithm of Cabello, the simple $O(n \log n)$-time algorithm in Theorem 1 yields an $O(n^2)$-time algorithm with ratio $0.4487$, and the LP-based algorithm in Theorem 2 yields a polynomial-time algorithm with ratio $0.4674$, for dispersion in $n$ (not necessarily disjoint) unit disks.*

It is likely that our method for proving Theorem 2, which uses projections for approximating distances, and linear programming for optimization, is also applicable to other optimization problems involving distances.

**Related work.** The dispersion problem in disks we study here is related to a few other problems in computational geometry. We mention several results that are more closely related to ours:

1. For labeling $n$ points with $n$ disjoint congruent disks, each point on the boundary of a distinct disk, such that radius of the disks is maximized, Jiang et al. [10] presented a $\frac{1}{2.98+\varepsilon}$-approximation algorithm, and proved that the problem is NP-hard to approximate with ratio more than $\frac{1}{1.0349}$.

2. For packing of $n$ axis-parallel congruent squares (congruent disks in the $L_\infty$ metric) in the same rectilinear polygon such that the side length of the squares is maximized, Baur and Fekete [1] presented a $\frac{2}{3}$-approximation algorithm, and proved that the problem is NP-hard to approximate with ratio more than $\frac{13}{14}$. A $\frac{2}{3}$-approximation algorithm for a related problem of packing $n$ unit disks in a rectangle without overlapping an existent set of $m$ unit disks in the same rectangle, has been obtained by Benkert et al. [2].

3. Given $n$ points in the plane, Demaine et al. [4] considered the problem of moving them to an *independent set* in the unit disk graph metric: that is, each point has to move to a position such that all pairwise distances are at least 1, and such that the maximum distance a point moved is minimized. They presented an approximation algorithm, which achieves a good ratio if the points are initially "far from" an independent set. However the approximation ratio becomes unbounded for instances that are "very close to" an independent set. Observe that in this problem, the optimum may be arbitrarily small, i.e., arbitrarily close to 0.

4. Dumitrescu and Jiang [6] obtained hardness results and approximation algorithms for two related geometric problems involving movement. The first is a constrained variant of the $k$-center problem, arising from a geometric client-server problem. The second is the problem of moving points towards an independent set, discussed previously.

## 2 A simple approximation algorithm for disjoint unit disks

In this section we present a very simple approximation algorithm **A1** with ratio $0.511$ for disjoint unit disks, and thereby prove Theorem 1.

The idea of the algorithm is as follows. Recall that $\delta$ is the minimum pairwise distance among the unit disks. Let $\sigma = \sigma(\delta)$ be a positive parameter to be specified; in particular, at the threshold distance $\delta = 2$ for disjoint unit disks, we have $\sigma(2) = 2.0883\ldots$, which is only slightly larger than $\delta$. Consider the *distance graph* of the unit disks for the parameter $\sigma$, which has a vertex for each disk, and an edge between two vertices if and only if the corresponding disks have distance at most $\sigma$. If there is a vertex of degree at least two in the distance graph, that is, if there is a disk close to two other disks, then a packing argument shows that the minimum pairwise distance of any three points in the three disks must be small. Thus simply placing the points at the disk centers already achieves a good approximation ratio. Otherwise, every vertex in the distance graph has degree at most one, and the edges form a matching. In this case, the disks that are close to each other are grouped into pairs. The distance between the two points in each pair can be slightly increased by moving them away from the disk centers, at the cost of possibly decreasing the distances between points in different pairs.

Let $\mathcal{D}$ be a set of $n$ (not necessarily disjoint) unit disks in the plane. The algorithm **A1** consists of three steps:

1. Compute the minimum pairwise distance $\delta$ of the disks in $\mathcal{D}$, and for each disk, find the two disks closest to it.

2. If the distance from some disk to its second closest disk is at most $\sigma = \sigma(\delta)$, return the $n$ disk centers as the set of points. Otherwise, proceed to the next step.

3. Place a point at the center of each disk. Then, for each disk, if the distance from the disk to its closest disk is at most $\sigma$, move the point away from the closest disk for a distance of $(\sigma - \delta)/4$, so that the two points in each close pair of disks are moved in opposite directions; we will show that $\delta < \sigma < \delta + 4$, thus the distance $(\sigma - \delta)/4$ is between 0 and 1, and each point remains in its own disk. Finally, return the set of points.

**Algorithm analysis.** The bottleneck for the running time of the algorithm **A1** is simply the computation of the two closest disks from each disk in step 1, which takes $O(n \log n)$ time [7, p. 306]. The other two steps of the algorithm can clearly be done in $O(n)$ time. For the proof of the approximation ratio, define the following function $f(s)$ for $s \geq 0$:

$$f(s) = \sqrt{(1+s)^2 + 1/2 + \sqrt{3(1+s)^2 - 3/4}}. \tag{2}$$

The function $f(\cdot)$ is increasing and $f(0) = \sqrt{3}$. The justification for step 2 of the algorithm **A1** is the following packing lemma. Here the disk with center $O$ is close to two other disks with centers $P$ and $Q$, respectively; see Figure 1.

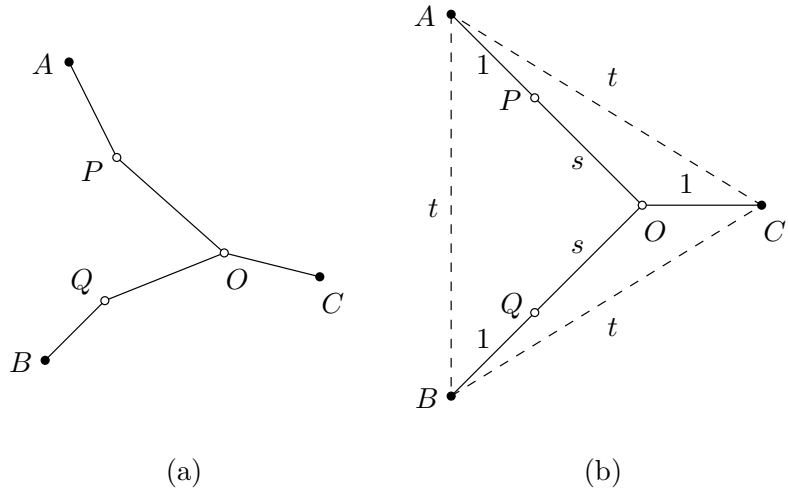

(a)          (b)

Figure 1: (a) A linkage of the five segments $AP, BQ, CO, OP, OQ$ for three points $A, B, C$ in three unit disks with centers $P, Q, O$, respectively. (b) The extreme configuration: $A, P, O$ are collinear, $B, Q, O$ are collinear, $|AP| = |BQ| = |CO| = 1$, $|OP| = |OQ| = s$, $|AC| = |BC| = |AB| = t$.

**Lemma 1.** *Let $A, B, C$ be three points in three unit disks with centers $P, Q, O$, respectively. Let $s = \max\{|OP|, |OQ|\}$ and $t = \min\{|AC|, |BC|, |AB|\}$. Then $t \leq f(s)$.*

*Proof.* Refer to Figure 1(a) for a linkage of the five segments $AP, BQ, CO, OP, OQ$ under the length constraints $\max\{|OP|, |OQ|\} = s$ and $\max\{|AP|, |BQ|, |CO|\} \leq 1$. Then, by a continuous movement argument (with the point $O$ fixed), it follows that the minimum pairwise distance $t$ of the three points $A, B, C$ is maximized by the extreme configuration in Figure 1(b).

Now consider the extreme configuration in Figure 1(b). Let $\theta = \frac{1}{2}\angle AOB$. In $\triangle AOB$, we have

$$t = 2(1+s)\sin\theta \implies \sin\theta = \frac{t}{2(1+s)}.$$

In $\triangle AOC$, we have

$$t^2 = 1^2 + (1+s)^2 - 2(1+s)\cos(\pi - \theta) \implies \cos\theta = \frac{t^2 - (1+s)^2 - 1}{2(1+s)}.$$

From $\sin^2\theta + \cos^2\theta = 1$, it follows that

$$t^2 + \left(t^2 - (1+s)^2 - 1\right)^2 = 4(1+s)^2.$$

Substitute $x = (1+s)^2$ and $y = t^2$, and we have

$$y + (y - x - 1)^2 = 4x,$$

which solves to

$$y + y^2 + x^2 + 1 - 2xy - 2y + 2x = 4x$$
$$y^2 - (2x+1)y + x^2 - 2x + 1 = 0$$
$$y = x + 1/2 \pm \sqrt{3x - 3/4}$$

Both solutions of $y$ are valid, but since we are deriving an upper bound for $t$, we take the larger solution

$$y = x + 1/2 + \sqrt{3x - 3/4}.$$

Thus, in this extreme case, we have

$$t^2 = (1+s)^2 + 1/2 + \sqrt{3(1+s)^2 - 3/4} = f^2(s).$$

It follows that, in general, $t \leq f(s)$. $\square$

Consider the following equation in $\sigma$:

$$\frac{\delta}{f(\sigma)} = \frac{\sigma + \delta}{2(\delta + 2)}. \tag{3}$$

The next lemma confirms that $\sigma$ exists and lies in the desired range:

**Lemma 2.** *There is a unique solution $\sigma$ to (3). Moreover, $\delta < \sigma < \delta + 4$.*

*Proof.* We first show that, for any $s \geq 0$, $s + 1 < f(s) < s + 2$:

$$f(s) = \sqrt{(1+s)^2 + 1/2 + \sqrt{3(1+s)^2 - 3/4}} > \sqrt{(1+s)^2} = s + 1,$$

$$f(s) = \sqrt{(1+s)^2 + 1/2 + \sqrt{3(1+s)^2 - 3/4}} < \sqrt{(1+s)^2 + 1 + \sqrt{4(1+s)^2}} = s + 2.$$

Since the function $f(s)$ is increasing in $s$ for $s \geq 0$, the left-hand side of (3) is decreasing in $\sigma$. On the other hand, the right-hand side of (3) is increasing in $\sigma$. If $\sigma \leq \delta$, then we would have the inequality

$$\frac{\delta}{f(\sigma)} \geq \frac{\delta}{f(\delta)} > \frac{\delta}{\delta + 2} = \frac{\delta + \delta}{2(\delta + 2)} \geq \frac{\sigma + \delta}{2(\delta + 2)}.$$

Similarly, if $\sigma \geq \delta + 4$, then we would have the inequality

$$\frac{\delta}{f(\sigma)} \leq \frac{\delta}{f(\delta + 4)} < \frac{\delta}{(\delta + 4) + 1} < 1 = \frac{(\delta + 4) + \delta}{2(\delta + 2)} \leq \frac{\sigma + \delta}{2(\delta + 2)}.$$

Therefore, there is a unique solution $\sigma$ to (3), and $\delta < \sigma < \delta + 4$. □

We now analyze the approximation ratio of the algorithm **A1**. Let ALG be the minimum pairwise distance of the points returned by the algorithm. Let OPT be the minimum pairwise distance of the optimal set of points. Let

$$c = c(\delta) = \frac{\delta}{f(\sigma)} = \frac{\sigma + \delta}{2(\delta + 2)}. \tag{4}$$

We next prove that the approximation ratio of the algorithm **A1** is at least $c$, namely that $\text{ALG} \geq c \cdot \text{OPT}$, by considering two cases:

- If the algorithm returns the $n$ disk centers as the set of points in step 2, then there is a disk such that the distances from the disk to its two closest disks are at most $\sigma$. By Lemma 1, we have $\text{OPT} \leq f(\sigma)$. Since $\text{ALG} = \delta$, it follows that

$$\frac{\text{ALG}}{\text{OPT}} \geq \frac{\delta}{f(\sigma)}. \tag{5}$$

- If the algorithm proceeds to step 3, then the distance from each disk to its second closest disk is more than $\sigma$. If two disks have distance at most $\sigma$, then they must be the closest disks of each other, and the movements of points in step 3 ensure that their two points have distance at least $\delta + 2(\sigma - \delta)/4 = (\sigma + \delta)/2$. On the other hand, if two disks have distance more than $\sigma$, then after the movements their two points have distance at least $\sigma - 2(\sigma - \delta)/4 = (\sigma + \delta)/2$. Thus $\text{ALG} \geq (\sigma + \delta)/2$. Since $\text{OPT} \leq \delta + 2$, it follows that

$$\frac{\text{ALG}}{\text{OPT}} \geq \frac{\sigma + \delta}{2(\delta + 2)}. \tag{6}$$

By (4), (5), and (6), the algorithm **A1** achieves an approximation ratio of $c(\delta)$ for $\delta \geq 0$. It can be verified that $c(\delta)$ is an increasing function of $\delta$ for $\delta \geq 0$. Thus, for dispersion in disjoint unit disks, the approximation ratio is

$$c(\delta) \geq c(2) = 0.5110\ldots, \quad \text{for } \delta \geq 2.$$

This completes the proof of Theorem 1.

# 3 An LP-based approximation algorithm for disjoint disks

**Warm-up: one dimension.** As a warm-up exercise, we first study the dispersion problem on the line and on a closed curve. In these two settings the problem can be solved exactly in polynomial time.

**Proposition 1.** *There exists a polynomial-time exact algorithm based on linear-programming for* DISPERSION *in disjoint intervals on the line and on a closed curve.*

*Proof.* For the line, the input consists on $n$ interior-disjoint intervals $[a_1, b_1], \ldots, [a_n, b_n]$, where (i) $a_i \leq b_i$, for $i = 1, \ldots, n$, and (ii) $b_i \leq a_{i+1}$, for $i = 1, \ldots, n-1$. Computing an optimal solution $x_i \in [a_i, b_i]$, $i = 1, \ldots, n$, amounts to solving the following linear program with the $n$ variables $x_i$ and $3n - 1$ constraints:

$$\begin{aligned}
\text{maximize} \quad & z & \text{(LP1)}\\
\text{subject to} \quad & \begin{cases} a_i \leq x_i \leq b_i & 1 \leq i \leq n \\ x_{i+1} - x_i \geq z, & 1 \leq i \leq n-1 \end{cases}
\end{aligned}$$

For a closed curve of length $L$, the input consists on $n$ interior-disjoint intervals $[a_1, b_1], \ldots, [a_n, b_n]$, where (i) $a_i \leq b_i$, for $i = 1, \ldots, n$, (ii) $b_i \leq a_{i+1}$, for $i = 1, \ldots, n-1$, and (iii) $0 \leq a_1$, $b_n \leq L$. Computing an optimal solution $x_i \in [a_i, b_i]$, $i = 1, \ldots, n$, amounts to solving the following linear program with the $n$ variables $x_i$ and $3n$ constraints:

$$\begin{aligned}
\text{maximize} \quad & z & \text{(LP2)}\\
\text{subject to} \quad & \begin{cases} a_i \leq x_i \leq b_i & 1 \leq i \leq n \\ x_{i+1} - x_i \geq z, & 1 \leq i \leq n-1 \\ x_1 + L - x_n \geq z \end{cases}
\end{aligned}$$

□

**Two dimensions.** Next we present and analyze the approximation algorithm **A2** and thereby prove Theorem 2. We first introduce some definitions and notations. Let $\Omega_1, \ldots, \Omega_n$ be $n$ pairwise disjoint disks of radii $r_1, \ldots, r_n$, and centers $o_1, \ldots, o_n$. Let $0 \leq \lambda < \lambda' < 1$ be two parameters. We set $\lambda = 1/2$ and $\lambda' = 3/4$ (with foresight) in order to maximize the approximation ratio. For $i = 1, \ldots, n$, let $\omega_i$ and $\omega_i'$ be two disks of radii $\lambda \cdot r_i$ and $\lambda' \cdot r_i$, respectively, that are concentric with $\Omega_i$. Note that $\omega_i \subset \omega_i' \subset \Omega_i$. Let $\alpha_{ij} \in [-\pi/2, \pi/2)$ be the direction (or angle) of the line determined by $o_i$ and $o_j$. For $\alpha \in [-\pi/2, \pi/2)$, let $\ell_\alpha$ be any line of direction $\alpha$. For two vectors $\overline{u} = (u_1, u_2)$, and $\overline{v} = (v_1, v_2)$, their dot product is $\langle \overline{u}, \overline{v} \rangle = u_1 v_1 + u_2 v_2$. The scalar projection of $\overline{v}$ onto $\overline{u}$ is the length of the orthogonal projection of the vector $\overline{v}$ onto $\overline{u}$, with a minus sign if the direction is opposite. It is given by the formula

$$\text{proj}_{\overline{u}} \overline{v} = \frac{\langle \overline{u}, \overline{v} \rangle}{|\overline{u}|}. \tag{7}$$

For two points, $p$ and $q$, let $\text{proj}_\alpha(p, q)$ denote the length of the projection of the segment $pq$ onto a line $\ell_\alpha$ of direction $\alpha$, i.e., onto the vector $(\cos \alpha, \sin \alpha)$. For simplicity, we write $\text{proj}_{ij}(q_i, q_j)$ instead of $\text{proj}_{\alpha_{ij}}(q_i, q_j)$.

The idea of the algorithm is as follows: Suppose we restrict the feasible region of each point $p_i$ from the given disk $\Omega_i$ to the smaller concentric disk $\omega_i$ of radius $\lambda \cdot r_i$. We then show the existence of a good approximation for the dispersion problem constrained to the smaller size disks. First, observe that the centers of the original disks $\Omega_i$ are still in the feasible regions for each of the $n$ points. So the $\frac{1}{2}$-approximation that we could easily achieve earlier, is still attainable. For instance, setting $\lambda = 0$ yields the algorithm CENTERS discussed earlier. Second, observe that if $\lambda$ is sufficiently small, then the distance between two points (in two smaller disks) can be well approximated by the projection of the segment connecting the two points onto the line connecting the centers of the two disks. Enclose each smaller disk $\omega_i$ in a suitable convex polygon $Q_i$, where $\omega_i \subset Q_i \subset \omega'_i \subset \Omega_i$. The length of each such projection can be expressed as a linear combination of the coordinates of the two points, and we can use linear programming to maximize the smallest projection length of an inter-point distance. All the constraints in the dispersion problem will be expressed as linear inequalities, at the cost of finding only an approximate solution. We now present the technical details.

We start with a technical lemma that will be used in establishing the approximation ratio of Algorithm **A2**.

**Lemma 3.** *Let $\lambda = 1/2$, $a \in [0, 1]$, and $\alpha \in [0, 2\pi)$. Then*

$$\frac{1 + a\lambda \cos \alpha}{\sqrt{1 + a^2 + 2a \cos \alpha}} \geq \frac{1}{\sqrt{2}}. \tag{8}$$

*Proof.* Since the numerator of the left hand side in (8) is non-negative, (8) is equivalent to

$$1 + a^2 \lambda^2 \cos^2 \alpha + 2a\lambda \cos \alpha \geq \lambda(1 + a^2 + 2a \cos \alpha). \tag{9}$$

After reducing the term $2a\lambda \cos \alpha$, (9) is equivalent to

$$1 + a^2 \lambda^2 \cos^2 \alpha \geq \lambda + \lambda a^2. \tag{10}$$

This follows easily from the following chain of inequalities

$$\lambda(1 + a^2) = \frac{1 + a^2}{2} \leq 1 \leq 1 + a^2 \lambda^2 \cos^2 \alpha, \tag{11}$$

as required. □

A key fact relating projections to distances is:

**Lemma 4.** *Let $\lambda = 1/2$. Consider two disjoint disks $\Omega_i$ and $\Omega_j$ at distance $\delta_{ij} = |o_i o_j|$. Let $p_i \in \Omega_i$ and $p_j \in \Omega_j$ be two points. Let $q_i \in \omega_i$ be the point on $o_i p_i$ at distance $\lambda|o_i p_i|$ from $o_i$. Similarly define $q_j \in \omega_j$ as the point on $o_j p_j$ at distance $\lambda|o_j p_j|$ from $o_j$. Then*

$$\frac{\text{proj}_{ij}(q_i, q_j)}{|p_i p_j|} \geq \frac{1}{\sqrt{2}}. \tag{12}$$

*Proof.* We can assume w.l.o.g. that $o_i = (0, 0)$ and $o_j = (1, 0)$, so that $\delta_{ij} = 1$. To represent points, we use complex numbers in the proof. Let Re$(z)$ denote the real part of a complex number $z \in \mathbb{C}$. The point $p_i$ is represented by $z_i$, where $z_i \in \mathbb{C}$, with $|z_i| \leq r_i$; hence $q_i$ is represented by $\lambda z_i$. Since the disks $\Omega_i$ and $\Omega_j$ are disjoint, we have $r_i + r_j \leq 1$. The point $p_j$ is represented by $1 + z_j$, where $z_j \in \mathbb{C}$, with $|z_j| \leq r_j$; hence $q_j$ is represented by $1 + \lambda z_j$. Write $z = z_j - z_i$, and note that $|z| \leq |z_i| + |z_j| \leq r_i + r_j \leq 1$. Let $z = a(\cos \alpha + i \sin \alpha)$ be the complex number representation of $z$, where $0 \leq a \leq 1$, and $\alpha \in [0, 2\pi)$. The segments $q_i q_j$ and $p_i p_j$ are then represented by $1 + \lambda(z_j - z_i) = 1 + \lambda z$ and $1 + z$, respectively.

Note that
$$\text{proj}_{ij}(q_i, q_j) = \text{Re}(1 + \lambda z) = 1 + a\lambda \cos \alpha,$$

and
$$|p_i p_j| = |1 + z| = |1 + a(\cos \alpha + \mathrm{i} \sin \alpha)| = \sqrt{1 + a^2 + 2a \cos \alpha}\ .$$

Hence by Lemma 3 we have
$$\frac{\text{proj}_{ij}(q_i, q_j)}{|p_i p_j|} = \frac{1 + a\lambda \cos \alpha}{\sqrt{1 + a^2 + 2a \cos \alpha}} \geq \frac{1}{\sqrt{2}},$$

as required. □

We now outline our LP-based algorithm **A2** for disjoint disks. Conveniently select disjoint convex polygons $Q_i$, $i = 1, \ldots, n$, such that $\omega_i \subset Q_i \subset \omega'_i \subset \Omega_i$, for each $i = 1, \ldots, n$. For instance let $Q_i$ be an axis-aligned square of side length $r_i$ concentric with $\omega_i$. Since $\Omega_i$ are pairwise disjoint, $Q_i$ are also pairwise disjoint. Moreover, since $Q_i$ and $Q_j$ are separated by the perpendicular bisector of $o_i o_j$, for any $q_i \in Q_i$ and $q_j \in Q_j$, the dot product $\langle \overline{q_i q_j}, \overline{o_i o_j} \rangle$ we use in formulating the LP is non-negative. We are lead to the following linear program, LP3, with the constraints expressed symbolically at this point. A set $\{q_1, \ldots, q_n\}$ of $n$ points is sought, where $q_i = (x_i, y_i) \in Q_i$, for $i = 1, \ldots, n$. LP3 maximizes the minimum pairwise projection on the line connecting the corresponding centers of the disks; that is, for each pair $(i, j)$, the length of the projection of the segment connecting the two points $q_i$ and $q_j$, on the line connecting the corresponding disk centers $o_i$ and $o_j$.

$$\begin{aligned}
\text{maximize} \quad & z \\
\text{subject to} \quad & \begin{cases} q_i \in Q_i, & 1 \leq i \leq n \\ \text{proj}_{ij}(q_i, q_j) \geq z, & 1 \leq i < j \leq n \end{cases}
\end{aligned} \tag{LP3}$$

**Writing the linear constraints.** Each symbolic constraint $q_i \in Q_i$ is implemented as four linear inequalities, one for each side of $Q_i$. Implement each symbolic constraint $\text{proj}_{ij}(q_i, q_j) \geq z$ as follows: Let $o_i = (\xi_i, \nu_i)$ be coordinates of $o_i$, for $i = 1, \ldots, n$ (part of the input). Consider a pair $(i, j)$, where $i < j$. Recall that $\alpha_{ij} \in [-\pi/2, \pi/2)$ is the angle of the line determined by $o_i$ and $o_j$. Assuming that $\xi_1 \leq \xi_2 \leq \ldots \leq \xi_n$, we have

$$\cos \alpha_{ij} = \frac{\xi_j - \xi_i}{|o_i o_j|}, \quad \sin \alpha_{ij} = \frac{\nu_j - \nu_i}{|o_i o_j|}. \tag{13}$$

Let $\overline{a_{ij}} = (\cos \alpha_{ij}, \sin \alpha_{ij})$, so that $|\overline{a_{ij}}| = 1$. Let $\overline{s_{ij}} = (x_j - x_i, y_j - y_i)$. According to (7),

$$\text{proj}_{ij}(q_i, q_j) = \frac{\langle \overline{a_{ij}} \cdot \overline{s_{ij}} \rangle}{|\overline{a_{ij}}|} = \langle \overline{a_{ij}} \cdot \overline{s_{ij}} \rangle = (x_j - x_i) \cos \alpha_{ij} + (y_j - y_i) \sin \alpha_{ij}.$$

As noted earlier the above expression for the projection is always non-negative. Consequently, for each pair $(i, j)$, where $i < j$, generate the constraint:

$$(x_j - x_i) \cos \alpha_{ij} + (y_j - y_i) \sin \alpha_{ij} \geq z,$$

where $\cos \alpha_{ij}$ and $\sin \alpha_{ij}$ are as in (13).

**Solving the linear program.**

**Lemma 5.** *For any given $\varepsilon > 0$, a $(1-\varepsilon)$-approximation of the solution of LP3 can be obtained in polynomial time.*

*Proof.* Recall that the linear program LP3 finds a point set $\{q_i = (x_i, y_i), i = 1, \ldots, n\}$, $q_i \in Q_i \subset \Omega_i$, for which the minimum projection is maximized. The constraints of the linear program LP3 involve irrational numbers, and hence it cannot be claimed that the original LP is solvable in polynomial time. However, it is enough to solve the LP up to some precision. For this, it is enough to approximate the numbers involved in the constraints up to some precision, which is polynomial in the error of the output [12]. Consequently, we can encode each coefficient into a rational number with $(1/\varepsilon)^{O(1)}$ bits. Then, for a constant $\varepsilon$, each coefficient has a constant number of bits, and the LP algorithm runs in polynomial time; e.g., $O(n^4)$ or $O(n^{3.5})$ using interior point methods. □

**Establishing the approximation ratio.**

**Lemma 6.** *For any given $\varepsilon > 0$, the approximation algorithm **A2** can achieve a ratio at least $\frac{1-\varepsilon}{\sqrt{2}}$ for pairwise disjoint disks.*

*Proof.* Let $\mathrm{OPT} = \zeta$ denote an optimal solution, given by $n$ points $p_1, \ldots, p_n$, where $p_i \in \Omega_i$, such that $|p_i p_j| \geq \zeta$, for all $i \neq j$, and $|p_i p_j| = \zeta$ for at least one pair $(i, j)$. Let $q_i \in \omega_i$ be the point on $o_i p_i$ at distance $\lambda |o_i p_i|$ from $o_i$. By Lemma 3 and 4, for all $i \neq j$,

$$\frac{\mathrm{proj}_{ij}(q_i, q_j)}{|p_i p_j|} = \frac{1 + a\lambda \cos\alpha}{\sqrt{1 + a^2 + 2a\cos\alpha}} \geq \frac{1}{\sqrt{2}}.$$

Since $|p_i p_j| \geq \zeta$, for all $i \neq j$, we immediately conclude that

$$\mathrm{proj}_{ij}(q_i, q_j) \geq \frac{\zeta}{\sqrt{2}}, \tag{14}$$

for all $i \neq j$.

Since $\omega_i \subset Q_i$, by Lemma 5, the point set found by the linear program satisfies $q_i \in Q_i \subset \Omega_i$, such that

$$\mathrm{proj}_{ij}(q_i, q_j) \geq \frac{1-\varepsilon}{\sqrt{2}} \cdot \zeta. \tag{15}$$

Since obviously, $|q_i q_j| \geq \mathrm{proj}_{ij}(q_i, q_j)$, we have found points $q_i \in Q_i \subset \Omega_i$, such that

$$|q_i q_j| \geq \frac{1-\varepsilon}{\sqrt{2}} \cdot \zeta = \frac{1-\varepsilon}{\sqrt{2}} \cdot \mathrm{OPT}.$$

Hence the approximation ratio is at least $\frac{1-\varepsilon}{\sqrt{2}}$, as claimed. For instance, when setting $\varepsilon = 10^{-4}$, we get a 0.707 approximation. □

**Reducing the number of constraints to $O(n)$.** Recall that $\mathrm{OPT} \leq 2\delta$, as shown in (1). Observe that the LP solution, $z^*$, is bounded from above as (recall that $\lambda' = 3/4$)

$$z^* \leq \delta + \frac{3(r_i + r_j)}{4} \leq \frac{7\delta}{4},$$

where $(i,j)$ are a closest pair of disks. It follows that there is no need to write any constraints for pairs of disks at distance larger than $7\delta$. Indeed, if now $(i,j)$ is such a pair, the distance between two points, one in $Q_i$ and one in $Q_j$, is at least

$$\delta_{ij} - \frac{3(r_i + r_j)}{4} \geq \delta_{ij} - \frac{3\delta_{ij}}{4} = \frac{\delta_{ij}}{4} > \frac{7\delta}{4} > z^*.$$

An easy packing argument shows that the number of pairs of disks at distance at most $7\delta$ is only $O(n)$: this is the same as the number of pairs of points at distance at most 7 times the minimum pairwise distance among $n$ points.

**Extension to any (fixed) dimension** $d$. We briefly sketch the differences. The balls $\omega_i$ and $\omega'_i$ are two smaller balls of radii $\lambda \cdot r_i$ and $\lambda' \cdot r_i$ concentric with $\Omega_i$, where $\lambda = 1/2$ and $\lambda' = 3/4$. $Q_i$ is any suitable convex polytope in $\mathbb{R}^d$ such that $\omega_i \subset Q_i \subset \omega'_i \subset \Omega_i$. The proof of the following lemma we need, closely mimics the proof of Lemma 4.

**Lemma 7.** *Let $\lambda = 1/2$. Consider two disjoint balls $\Omega_i$ and $\Omega_j$ at distance $\delta_{ij} = |o_i o_j|$. Let $p_i \in \Omega_i$ and $p_j \in \Omega_j$ be two points. Let $q_i \in \omega_i$ be the point on $o_i p_i$ at distance $\lambda |o_i p_i|$ from $o_i$. Similarly define $q_j \in \omega_j$ as the point on $o_j p_j$ at distance $\lambda |o_j p_j|$ from $o_j$. Then*

$$\frac{\mathrm{proj}_{ij}(q_i, q_j)}{|p_i p_j|} \geq \frac{1}{\sqrt{2}}. \tag{16}$$

*Proof.* We can assume w.l.o.g. that $o_i = (0, \ldots, 0)$ and $o_j = (1, 0, \ldots, 0)$. To represent points, we use vectors in the proof. Let $\overline{e_1} = \overline{(1, 0, \ldots, 0)}$. Note that $|\overline{e_1}| = 1$. The point $p_i$ is represented by $\overline{v_i}$, where $|\overline{v_i}| \leq r_i$; hence $q_i$ is represented by $\lambda \overline{v_i}$. Since the two balls $\Omega_i$ and $\Omega_j$ are disjoint, we have $r_i + r_j \leq 1$. The point $p_j$ is represented by $\overline{e_1} + \overline{v_j}$, where $|\overline{v_j}| \leq r_j$; hence $q_j$ is represented by $\overline{e_1} + \lambda \overline{v_j}$. Write $\overline{v} = \overline{v_j} - \overline{v_i}$, and note that $|\overline{v}| \leq |\overline{v_i}| + |\overline{v_j}| \leq r_i + r_j \leq 1$. Let $a = |\overline{v}|$; clearly $0 \leq a \leq 1$. Let $\alpha \in [0, 2\pi)$ be the angle made by $\overline{v}$ with $\overline{e_1}$. We have

$$|p_i p_j| = |\overline{e_1} + \overline{v_j} - \overline{v_i}| = |\overline{e_1} + \overline{v}|.$$
$$|q_i q_j| = |\overline{e_1} + \lambda \overline{v_j} - \lambda \overline{v_i}| = |\overline{e_1} + \lambda \overline{v}|.$$

By (7), we have

$$\mathrm{proj}_{ij}(q_i, q_j) = \mathrm{proj}_{\overline{e_1}}(\overline{e_1} + \lambda \overline{v}) = \frac{\langle \overline{e_1}, \overline{e_1} + \lambda \overline{v} \rangle}{|\overline{e_1}|} = \langle \overline{e_1}, \overline{e_1} + \lambda \overline{v} \rangle$$
$$= \langle \overline{e_1}, \overline{e_1} \rangle + \langle \overline{e_1}, \lambda \overline{v} \rangle = 1 + \lambda \langle \overline{e_1}, \overline{v} \rangle = 1 + a\lambda \cos \alpha,$$

that is, the same expression as in the planar case. Further, by the cosine formula we have

$$|p_i p_j| = |\overline{e_1} + \overline{v}| = \sqrt{|\overline{e_1}|^2 + |\overline{v}|^2 + 2|\overline{e_1}||\overline{v}| \cos \alpha} = \sqrt{1 + a^2 + 2a \cos \alpha},$$

matching again the expression from the planar case.

We conclude in the same way by using Lemma 3:

$$\frac{\mathrm{proj}_{ij}(q_i, q_j)}{|p_i p_j|} = \frac{1 + a\lambda \cos \alpha}{\sqrt{1 + a^2 + 2a \cos \alpha}} \geq \frac{1}{\sqrt{2}},$$

as required. $\square$

The symbolic constraints are implemented analogous to those for the planar case. There exists a convex polytope $Q \subset \mathbb{R}^d$ and a function $f(d)$ such that $\omega \subset Q \subset \omega' \subset \Omega$, where $Q$ has $f(d)$ facets, and $\omega$, $\omega'$ and $\Omega$ are concentric balls of radii $1/2$, $3/4$ and $1$, respectively (note that already for $d = 5$, a concentric unit hyper-cube is *not* contained in the unit ball!). The polytope $Q_i$ is a translate of $r_i Q$ placed at $o_i$, so that $\omega_i \subset Q_i \subset \omega'_i \subset \Omega_i$. Each symbolic constraint $q_i \in Q_i$ is implemented as $f(d)$ linear inequalities, one for each facet of $Q_i$. Each symbolic constraint $\text{proj}_{ij}(q_i, q_j) \geq z$ implements the dot products from the proof of Lemma 7. Since (again) there is no need to write any constraints for pairs of balls at distance larger than $7\delta$, and the number of such pairs is linear in $n$ for fixed $d$ [11, p. 211], the total number of constraints is $O(n)$. The approximation ratio remains the same as for the planar case, namely $\frac{1-\varepsilon}{\sqrt{2}}$, for any given $\varepsilon > 0$, e.g., $0.707$ for $\varepsilon = 10^{-4}$. This completes the proof of Theorem 2.

## 4 A hybrid algorithm for unit disks

In this section we prove Theorem 3. For dispersion in (not necessarily disjoint) unit disks, Cabello [3] presented a hybrid algorithm that applies two different algorithms PLACEMENT and CENTERS and then returns the better solution. We first briefly review Cabello's analysis for his hybrid algorithm, then present an improved hybrid algorithm that uses the algorithm PLACEMENT in combination with either the simple $O(n \log n)$-time algorithm in Theorem 1 or the LP-based algorithm in Theorem 2. In the following, let $\text{OPT} = 2x$ and $\delta = 2\mu$. We can assume w.l.o.g. that $\delta \leq 2$, as otherwise the unit disks are disjoint. We also record the obvious inequalities:

$$\delta \leq \text{OPT} \leq \delta + 2 \leq 4 \quad \Leftrightarrow \quad \mu \leq x \leq 1 + \mu \leq 2. \tag{17}$$

The algorithm PLACEMENT, which runs in $O(n^2)$ time, achieves a ratio of

$$c_1(x) = \frac{-\sqrt{3} + \sqrt{3}x + \sqrt{3 + 2x - x^2}}{4x}, \quad \text{for } 1 \leq x \leq 2, \tag{18}$$

and a ratio of at least $\frac{1}{2}$ for $0 \leq x \leq 1$.

The algorithm CENTERS achieves a ratio of

$$c_2(x) = \frac{x-1}{x}, \quad \text{for } x \geq 1, \tag{19}$$

which is at least $\frac{1}{2}$ for $x \geq 2$.

Since $c_1(x)$ is decreasing in $x$ and $c_2(x)$ is increasing in $x$, the minimum approximation ratio of the hybrid algorithm occurs at the intersection of the two curves $c_1(x)$ and $c_2(x)$ for $1 \leq x \leq 2$: more precisely, $c_1(x) = c_2(x) = \frac{1}{\sqrt{5-2\sqrt{3}}+1} = 0.4465\ldots(1/2.2393\ldots)$ for $x = 1 + \frac{1}{\sqrt{5-2\sqrt{3}}} = 1.8068\ldots$.

Recall the approximation ratio $c(\delta)$ of our algorithm **A1** as a function of the minimum pairwise distance $\delta$ of the disks; see Equation (4). Now define

$$a_1(x) = c(2x - 2), \quad \text{for } x \geq 1. \tag{20}$$

Then we have the following lemma:

**Lemma 8.** *For any $x \geq 1$, the algorithm **A1** achieves an approximation ratio at least $a_1(x)$ for dispersion in unit disks.*

*Proof.* Recall that the function $c(\delta)$ is increasing in $\delta$. From (17) we deduce that $\delta \geq \text{OPT} - 2 = 2(x-1)$. Thus our algorithm **A1** achieves an approximation ratio of at least $c(\delta) \geq c(2x-2) = a_1(x)$ for $x \geq 1$. □

It can be verified that for $x \geq 1$, $a_1(x)$ is an increasing functions of $x$. If we replace the algorithm CENTERS by our algorithm **A1** in the hybrid algorithm, then the two curves $c_1(x)$ and $a_1(x)$ intersect at $x = 1.7750\ldots$ and, correspondingly, the approximation ratio of the new hybrid algorithm is at least $0.4487\ldots(1/2.2284\ldots)$.

We next discuss the hybrid algorithm that runs PLACEMENT and **A2**. Obviously the $n$ disks of radius $\mu \leq 1$ concentric with the $n$ input unit disks are pairwise-disjoint. The hybrid algorithm runs PLACEMENT on the given unit disks and **A2** on the disks of radius $\mu$ and then returns the better solution. Clearly the solution is valid, and it remains to analyze the approximation ratio. We will show that for any given $\varepsilon > 0$, it can achieve a ratio at least $\frac{(1-\varepsilon)\sqrt{2}}{1+\sqrt{9-2\sqrt{6}}} = (1-\varepsilon) \cdot 0.46749\ldots$. By taking then $\varepsilon = 10^{-4}$, we get a $0.4674$-approximation.

We start by relating the optimal solution OPT for the unit disks to the optimal solution $\text{OPT}_\mu$ for the smaller disjoint disks.

**Lemma 9.** *For a problem instance with $\mu \in [0, 1]$, we have $\text{OPT}_\mu \geq \text{OPT} - 2(1 - \mu)$.*

*Proof.* Consider an optimal solution given by $n$ points $p_1, \ldots, p_n$, where $p_i \in \Omega_i$, such that $|p_i p_j| \geq \text{OPT}$, for all $i \neq j$, and $|p_i p_j| = \text{OPT}$ for at least one pair $(i, j)$. Let $q_i \in \Omega_i$ be the point on $o_i p_i$ at distance $\mu |o_i p_i|$ from $o_i$. Obviously, the set $\{q_i : i = 1, \ldots, n\}$ is a valid solution for dispersion in the disks of radius $\mu$ concentric with the unit disks $\Omega_1, \ldots, \Omega_n$. Moreover, since $|p_i q_i| \leq 1 - \mu$, for $i = 1, \ldots, n$, by the triangle inequality we have $|q_i q_j| \geq \text{OPT} - 2(1-\mu)$, for all $i \neq j$. Consequently, $\text{OPT}_\mu \geq \text{OPT} - 2(1-\mu)$, as claimed. □

Now define
$$a_2(x, \mu) = \frac{x - 1 + \mu}{x\sqrt{2}}, \quad \text{for } x \in [1, 2], \text{ and } \mu \in [0, 1]. \tag{21}$$

Observe that $a_2(x, \mu)$ is an increasing function in both arguments. Then we have the following analogous lemma for our algorithm **A2**.

**Lemma 10.** *Consider a problem instance with $\mu \in [0, 1]$ and $x \in [1, 1 + \mu]$ as in (17). For any given $\varepsilon > 0$, the algorithm **A2** can achieve an approximation ratio at least $(1-\varepsilon) \cdot a_2(x, \mu)$ for dispersion in unit disks.*

*Proof.* By Lemma 9, we have $\text{OPT}_\mu \geq \text{OPT} - 2(1-\mu)$. By Lemma 6, for any given $\varepsilon > 0$, the algorithm **A2** can achieve an approximation ratio at least $\frac{1-\varepsilon}{\sqrt{2}}$, and therefore find a point set in the disks of radius $\mu$ with minimum inter-point distance at least
$$\frac{1-\varepsilon}{\sqrt{2}} \cdot \text{OPT}_\mu \geq \frac{1-\varepsilon}{\sqrt{2}} \cdot (\text{OPT} - 2(1-\mu)) = \frac{1-\varepsilon}{\sqrt{2}} \cdot (2x - 2(1-\mu)).$$

Equivalently, the approximation ratio obtained is at least
$$\frac{1-\varepsilon}{\sqrt{2}} \cdot \frac{2x - 2(1-\mu)}{2x} = (1-\varepsilon) \cdot \frac{x-1+\mu}{x\sqrt{2}} = (1-\varepsilon) \cdot a_2(x, \mu).$$
□

For a fixed $\mu \in [0,1]$, let $x_1(\mu)$ be the solution in $[1, \infty)$ of the equation $c_1(x) = a_2(x, \mu)$, or

$$\frac{-\sqrt{3} + \sqrt{3}x + \sqrt{3 + 2x - x^2}}{4x} = \frac{x - 1 + \mu}{x\sqrt{2}}. \tag{22}$$

For solving (22), make the substitution $y = x - 1$, and get

$$\sqrt{3}y + \sqrt{4 - y^2} = 2\sqrt{2}(y + \mu)$$
$$(3 - \sqrt{6})y^2 + (4 - \sqrt{6})\mu y + (2\mu^2 - 1) = 0$$
$$y_{1,2} = \frac{-(4 - \sqrt{6})\mu \pm \sqrt{12 - 4\sqrt{6} - 2\mu^2}}{2(3 - \sqrt{6})}.$$

Since the solution with a minus sign is negative, hence infeasible, we get

$$y_1(\mu) = \frac{-(4 - \sqrt{6})\mu + \sqrt{12 - 4\sqrt{6} - 2\mu^2}}{2(3 - \sqrt{6})}, \quad x_1(\mu) = 1 + y_1(\mu). \tag{23}$$

Observe that $y_1(\mu)$ and $x_1(\mu)$ are both decreasing functions of $\mu$ for $0 \le \mu \le 1$, with $y_1(0) = \frac{1}{\sqrt{3-\sqrt{6}}} = 1.34\ldots$, and $y_1(1/\sqrt{2}) = 0$.

Let $\mu_0$ be the solution in $[0,1]$ of the equation $y_1(\mu) = \mu$, or

$$\frac{-(4 - \sqrt{6})\mu + \sqrt{12 - 4\sqrt{6} - 2\mu^2}}{2(3 - \sqrt{6})} = \mu. \tag{24}$$

After rearranging terms and squaring, we get

$$\mu_0 = \sqrt{\frac{9 + 2\sqrt{6}}{57}} = \frac{1}{\sqrt{9 - 2\sqrt{6}}} = 0.4938\ldots. \tag{25}$$

**Lemma 11.**
$$c_1(1 + \mu_0) = a_2(1 + \mu_0, \mu_0) = \frac{\sqrt{2}}{1 + \sqrt{9 - 2\sqrt{6}}} = 0.4674\ldots. \tag{26}$$

*Proof.* By definitions we have

$$c_1(1 + \mu_0) = c_1(1 + y_1(\mu_0)) = a_2(1 + y_1(\mu_0), \mu_0) = a_2(1 + \mu_0, \mu_0)$$
$$= \frac{(1 + \mu_0) - 1 + \mu_0}{\sqrt{2}(\mu_0 + 1)} = \frac{\sqrt{2}\mu_0}{\mu_0 + 1} = \frac{\sqrt{2}}{1 + \frac{1}{\mu_0}} = \frac{\sqrt{2}}{1 + \sqrt{9 - 2\sqrt{6}}} = 0.4674\ldots,$$

as required. □

It suffices to prove the following lemma and then apply Lemma 10 to complete the proof of the theorem.

**Lemma 12.** *For every $\mu \in [0,1]$ we have*

$$\min_{x \in [1, 1+\mu]} \max\{c_1(x), a_2(x, \mu)\} \ge \frac{\sqrt{2}}{1 + \sqrt{9 - 2\sqrt{6}}} = 0.4674\ldots. \tag{27}$$

*Proof.* Let $h(x,\mu) = \max\{c_1(x), a_2(x,\mu)\}$. We distinguish two cases.

*Case 1*: $\mu \leq \mu_0$. By the definition of $\mu_0$ we have $\mu_0 = y_1(\mu_0) \leq y_1(\mu)$. Observe that $1 \leq x \leq 1 + \mu \leq 1 + \mu_0$. By (26), and the monotonicity of $c_1(\cdot)$, we have

$$h(x,\mu) \geq c_1(x) \geq c_1(1+\mu) \geq c_1(1+\mu_0) = \frac{\sqrt{2}}{1+\sqrt{9-2\sqrt{6}}} = 0.4674\ldots.$$

*Case 2*: $\mu \geq \mu_0$. If $x \leq 1 + \mu_0$ then $1 \leq x \leq 1 + \mu_0 \leq 1 + \mu$. By (26) and the monotonicity of $c_1(\cdot)$,

$$h(x,\mu) \geq c_1(x) \geq c_1(1+\mu_0) = \frac{\sqrt{2}}{1+\sqrt{9-2\sqrt{6}}} = 0.4674\ldots.$$

If $x \geq 1 + \mu_0$ then $1 \leq 1 + \mu_0 \leq x \leq 1 + \mu$. By (26) and the monotonicity of $a_2(\cdot,\mu)$ and $a_2(x,\cdot)$,

$$h(x,\mu) \geq a_2(x,\mu) \geq a_2(1+\mu_0,\mu) \geq a_2(1+\mu_0,\mu_0) = \frac{\sqrt{2}}{1+\sqrt{9-2\sqrt{6}}} = 0.4674\ldots.$$

Hence the inequality (27) is satisfied in both cases. $\square$

This completes the proof of Theorem 3.

## 5 Summary

Recall our two algorithms **A1** and **A2** and the two algorithms PLACEMENT and CENTERS by Cabello [3]. In conclusion, we summarize the current best approximation ratios for the three variants of dispersion in disks:

- Arbitrary (not necessarily unit or disjoint): $3/8 = 0.375$ by PLACEMENT.

- Unit (not necessarily disjoint): $0.4674$ by **A2** plus PLACEMENT, which improves $0.4465$ by CENTERS plus PLACEMENT.

- Disjoint (not necessarily unit): $0.707$ by **A2**, which improves $0.5$ by CENTERS.

**Acknowledgment.** The authors would like to thank Andreas Björklund for pertinent comments on an earlier version of the manuscript.